# Why Jet Power and Star Formation Are Uncorrelated in Active Galaxies

David Garofalo [1], Brent McDaniel [1] and Max North [2]

[1] Department of Physics, Kennesaw State University, USA
[2] Department of Information Systems, Kennesaw State University, USA

**Abstract:** Jet luminosity from active galaxies and the rate of star formation have recently been found to be uncorrelated observationally. We show how to understand this in the context of a model in which powerful AGN jets enhance star formation for up to hundreds of millions of years while jet power decreases in time, followed by a longer phase in which star formation is suppressed but coupled to jet power increasing with time. We also highlight characteristic differences depending on environment richness in a way that is also compatible with the observed SEDs of high redshift radio galaxies. While the absence of a direct correlation between jet power and star formation rate emerges naturally, our framework allows us to also predict the environment richness, range of excitation and redshift values of radio AGN in the jet power-star formation rate plane.

## 1. Introduction

While some evidence for positive active galactic nuclei (AGN) feedback exists at low redshift in Seyfert galaxies (e.g. Heckmann et al 2004, Prajwel et al 2022), in quasars at high redshift (e.g. Cresci & Maiolino 2018; Kokorev et al 2024), and on star formation (e.g. Kalfountzou et al 2012; Zinn et al 2013; Kimmig et al 2025), the idea that powerful AGN feedback is mostly negative comes from the observation that massive elliptical galaxies are red and dead (Cattaneo et al 2009; Fabian 2012; Fabian et al 2024). While negative AGN feedback limits the rate of star formation and the size of galaxies, positive AGN feedback can also manifest itself on the stellar velocity dispersion. It is thought that negative feedback is needed to explain the observation that galaxies are limited in size (Silk & Rees 1998; Di Matteo et al 2005; Croston et al 2006; Nesvadba et al 2006; Nesvada et al 2010; Silk & Mamon 2012). The correct combination of positive and negative feedback from AGN must be capable of explaining the black hole scaling relations (e.g. Magorrian et al 1998; Ferrarese & Merritt 2000). For example, evidence has emerged recently that two separate M-$\sigma$ relations exist (Sahu, Graham & Davis 2019). Resolving the nature of AGN feedback in this context is thus a key question in the formation and evolution of galaxies.

Jin et al (2025) have explored 5578 radio AGN (and their kinetic, or jet mode, feedback) in the LOFAR Two-Metre Sky Survey (LoTSS) and the Mapping Nearby Galaxies at APO (MaNGA) survey and find that massive radio AGN host galaxies are mostly quiescent, but their degree of star formation suppression is uncorrelated with jet luminosity. We apply our paradigm for black hole accretion and jet formation (Garofalo, Evans & Sambruna 2010) to evaluate the relation between jet power and star formation rate as a way of shedding light on these observations from a theoretical perspective, showing why these quantities cannot be directly correlated. In Section 2 we describe the features of the model and extract the quantitative connection between jet power and star formation rate, explaining the distribution found in Jin et al (2025) and make predictions for the distribution of redshift, excitation level (i.e. abundance of emission lines), and richness factor (i.e. the number of galaxies within a specified region of space). We then compare with the timescale for star formation suppression determined in a sample of high-z radio galaxies. In Section 3 we conclude.

## 2. Star Formation as a Function of Jet Power from Theory

Singh et al (2021) have applied an analytic model for black holes (Garofalo, Evans & Sambruna 2010) to explain the distribution of AGN in the star formation rate – stellar mass (SFR-SM) plane of Comerford et al (2020). The basic idea is that powerful jetted AGN are produced when accretion settles into counter-rotation around a high spinning black hole following a merger. This counter-rotating phase is associated with the enhancement of star formation (e.g. Kalfountzou et al 2012; Singh et al 2021) because the jet pushes gas into a state of higher density (Singh et al 2024). But once the black hole spins down to zero, the Bardeen-Petterson effect (Bardeen & Petterson 1975) disappears and a new disk orientation is generated (Garofalo, Joshi et al 2020). Because jets tap into the rotational energy of a black hole, the jet is weak and disappears in the transition through zero spin but once the spin is high enough in corotation, a new jet is generated. Given the shift in disk orientation, the new jet direction differs from the previous one, and this means that the jet impinges directly onto the interstellar medium, heating it and eventually suppressing star formation. This is shown in our schematic in Figure 1.

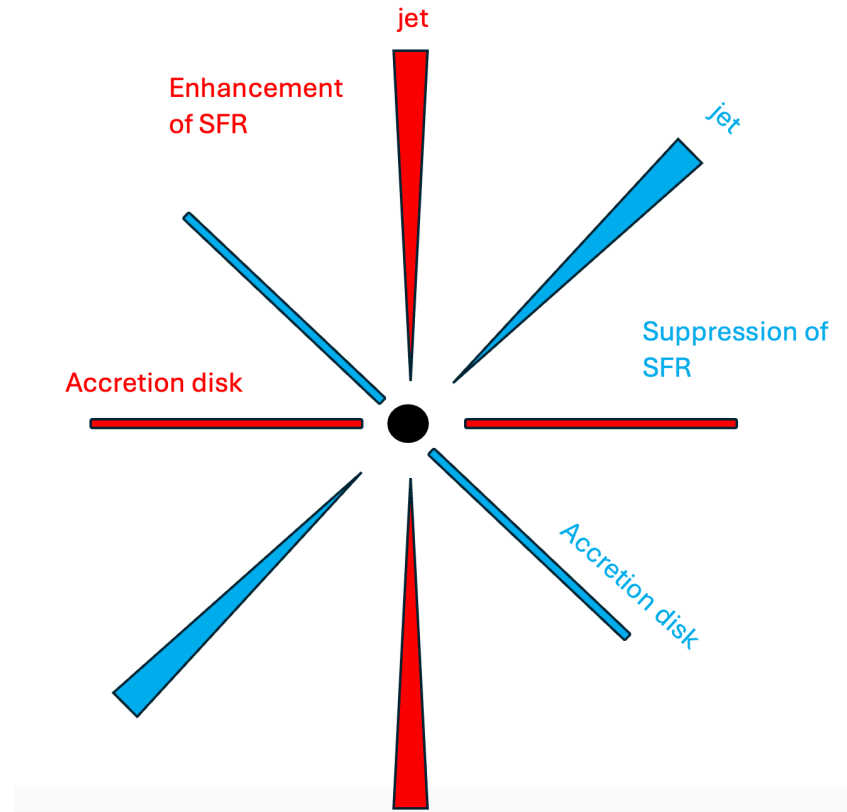

**Figure 1.** A counter-rotating black hole is triggered by a merger and a jet is produced that enhances the rate of star formation (red system – vertical jet). In the transition through zero black hole spin, a new disk orientation is generated and when a new jet is formed, it is tilted with respect to the previous orientation. This means the new jet suppresses the star formation rate by heating the gas (blue system – tilted jet). This effect becomes increasingly more relevant in richer environments.

These phases are described in Singh et al (2021) in their Figure 5 and our Figure 2 from which we extract the values shown in Table 1. The green paths in Figure 2 are counter-rotating phases while the pink paths are corotating phases. The former is associated with enhancement of star formation while the latter is associated with star formation suppression. In this work, we want to derive a jet power vs. star formation rate relation to address the data in Jin et al (2025). For this purpose, we need the theoretical expression for jet power (from Garofalo, Evans & Sambruna 2010) and its relation to the star formation rate which we can construct by matching values of jet power from theory to the turning points determined in Figure 2. To understand the details of Figure 2 in order to accomplish our goals, we imagine that a merger occurs and that this triggers star formation and eventually an active galaxy which places the system on the radio quiet AGN line because the jet that is also triggered in these systems has yet to have any impact. In isolated field environments this means an initial star formation rate (SFR) of 0.3 solar masses/yr. and a stellar mass value (SM) of $3.2 \times 10^{10}$ solar masses (see Figure 2). These values correspond to log SFR = -0.5 and log SM = 10.5. Given the division between FRII jets and FRI jets in Figure 2, we comment briefly on the FRI/FRII dichotomy. The main observational difference is that FRI jets tend to be core bright while FRII jets are edge bright. Due to the SFR-enhancement resulting from an FRII jet, the system evolves upward and to the right as stars form. The peak SFR is estimated to be 10 solar masses/yr. which gives it a log SFR = 1. Since the time during which the SFR is enhanced is given by the time it takes to spin down a counter-rotating BH at the Eddington rate (i.e. about $10^7$ yrs.), we can estimate the total number of stars formed during this enhancement phase. This adds just under an additional $10^8$ solar masses in stars giving it a log SM = 10.506. The end of the green line corresponds to BH spin down, after which, continued accretion spins the black hole up in corotation which corresponds to a period of negative feedback on SFR due to the disk wind and tilted jet that lasts about $10^8$ yrs. (see Singh et al 2021 for further details) that brings the SFR back to its original value prior to the disappearance of the jet. This adds an additional mass in stars of about 10 solar masses/yr. x $10^8$ yr. which is $10^9$ solar masses. Hence, the system ends up at an SM given by log SM = log ($3.2 \times 10^{10} + 10^8 + 10^9$) = 10.52. These values explain the range of SFR and SM for the green and pink paths for field environments. Our goal in this work is to generate theoretical values of SFR and jet power to show how the lack of correlation between them emerges. For this purpose, we need to be able to associate a jet power for all points of the green and pink paths. Theory has it that jet power depends on black hole mass and black hole spin, the latter being inferred by the fact that the transition from green to pink is where counter-rotation transitions to corotation. If one assumes that the system is born with high spin in counter-rotation and that it accretes near the Eddington limit, we can estimate values of BH spin for given SFR values. To determine the black hole mass along the green and pink paths, instead, we will use the black hole scaling relations that connect SM to BH mass (e.g. Gultekin et al 2009). Since SM changes little during the various phases, we adopt an average or characteristic SM value. For group and cluster environments, we follow the same procedure adopted above for field environments and obtain the values in rows 2 and 3 in Table 1. Note that the characteristic paths for denser environments have longer pink paths. This is due to effects that we briefly describe but do not go into in any detail (for these see Singh et al 2021). The characteristic BH mass is larger in these environments, which makes their characteristic jets more powerful. This leads to a larger SFR enhancement but also to a change in the accretion flow from thin and radiatively efficient to hot and advection dominated. This slows the evolution of the spin down and makes the slope on the SFR-SM plane drop for the green paths compared to the green path for field environments. As a result of this change in accretion, jets are no longer subject to jet suppression and the tilted jet in corotation can operate until the accretion fuel is consumed, and the active galaxy is dead. This is why the SFR in denser environments drops below the values in field environments. Because counter-rotation enhances the SFR while the black hole spin drops, and jet power increases with increase in black hole spin, we see qualitatively that an anti-correlation is instantiated during the green phase between jet power and SFR. This we want to make quantitative.

**Table 1.** Range of star formation rates during counter and co-rotating phases and characteristic stellar mass as a function of environment from Figure 2. While pink paths continue to drop over billions of years, for the purpose of calculations the values for the pink paths used in Table 1 are associated with a timescale of order 1 billion years.

| Environment | counter-rotating phase | corotating phase | stellar mass |
| --- | --- | --- | --- |
| Fields | $-0.5 < \log \text{SFR} < 1.0$ | $-0.5 < \log \text{SFR} < 1.0$ | $\log (M_\star / M_\odot) = 10.5$ |
| Groups | $0.5 < \log \text{SFR} < 1.6$ | $-2.1 < \log \text{SFR} < 1.6$ | $\log (M_\star / M_\odot) = 11.0$ |
| Clusters | $1.5 < \log \text{SFR} < 2.2$ | $-2.4 < \log \text{SFR} < 2.2$ | $\log (M_\star / M_\odot) = 11.5$ |

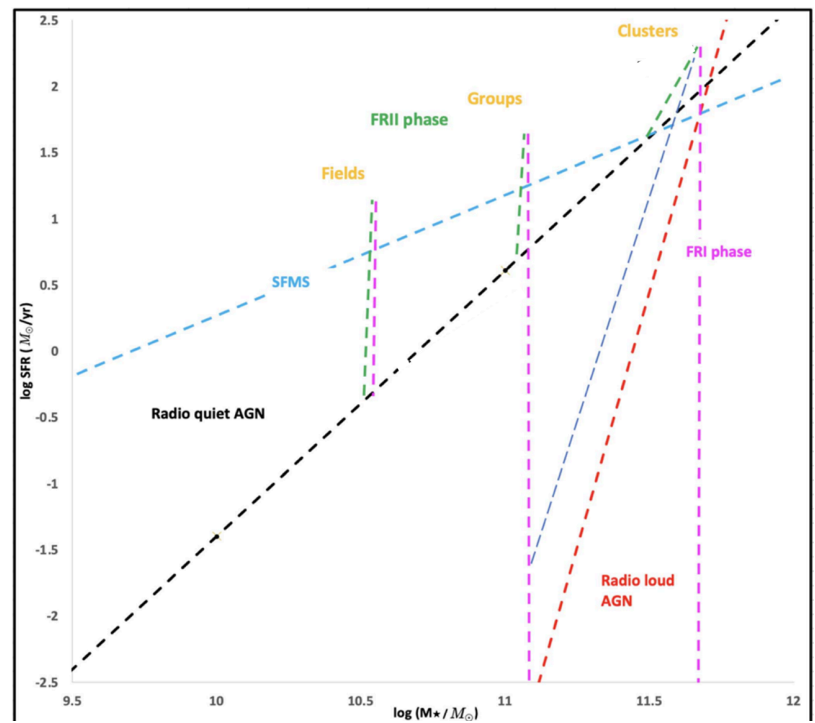

**Figure 2.** SFR vs. stellar mass from Singh et al 2021. Characteristic paths for isolated field, group and rich environments. A counter-rotating black hole is triggered in a merger which leads to a jet enhancement of the star formation rate (green paths). Once the transition through zero black hole spin occurs and a new disk orientation is generated, the new jet suppresses star formation. FRII jets are produced during green paths while FRI jets belong to pink paths. The location and slope of the radio loud AGN of Comerford et al 2020 are represented by the red dashed line.

With the values of BH spin and BH mass we can use the analytic model to derive jet power, which we then plot as a function of SFR. To accomplish this, we need the theoretical jet power expression from Garofalo, Evans & Sambruna (2010) as follows,

$$L_{jet} = 5 \times 10^{47} \text{ erg s}^{-1} \beta^2 (B_d/10^5 \text{ Gauss})^2 m_9^2 a^2 (1.5-a), \quad (1)$$

with

$$\beta = -1.5a^3 + 12a^2 - 10a + 7 - 2 \times 10^{-3}(a-0.65)^{-2} + 0.1(a+0.95)^{-1} + 2 \times 10^{-3}(a-0.055)^{-2}. \quad (2)$$

The dimensionless spin parameter *a* spans the range -1 < *a* < 1, with negative values representing counter-rotating accretion and positive values representing corotating accretion. $B_d$ is the magnetic field threading the inner accretion disk and $m_9$ is the mass of the black hole in terms of $10^9$ solar masses (i.e. if the mass of the black hole is 1 billion solar masses, $m_9$ = 1). The above expression is a fit from a numerical calculation and is designed to capture the numerical result, but it has singular points near which it cannot be used. While we use this expression because it provides a quantitative estimate of jet power, there are other jet power versus BH spin relations from numerical simulations that we could use just as well and that would provide the same qualitative conclusion to our study. The key to our result is not so much the actual jet power but the fact that such systems are triggered in counter-rotation and must spin down over time and lead to a tilted jet when the system spin up again in corotation. We could, in fact, use the jet power as a function of spin from general relativistic magnetohydrodynamic simulations (e.g. Tchekhovskoy, McKinney & Narayan, 2012 Figure 4; Lowell et al 2024, Figure 3), which display a slightly larger jet power (or jet efficiency) in counter-rotation compared to corotation as long as the spin is not larger than about 0.5 in corotation, which tends to be a limit for the accreting black holes in the densest environments. In order to evaluate $m_9$ from the SM values listed in the fourth column of Table 1, we appeal to the black hole mass-galaxy scaling relation connecting stellar mass to black hole mass (Reines & Volonteri 2015). We do not estimate magnetic fields so we will derive a jet power per $(B_d/10^5 \text{ Gauss})^{-2}$. Given the black hole mass-dependence on environment reported in Table 1, the jet power expressions will obviously be higher in clusters and lowest in fields per fixed black hole spin value.

Within our theoretical framework it is now possible to track both the jet power and the star formation rate as a function of time. To accomplish this, we take advantage of the existence of certain characteristic timescales. The first is the time to spin a high-spin BH down in counter-rotation at the Eddington limit. This takes about $10^7$ years. This timescale is respected for the average post-merger accreting BH in field and group environments in the paradigm. We therefore know the time it takes for the SFR to reach its peak value for the green paths in both field and group environments in Figure 2. This timescale also allows us to estimate the time to spin the BH down to about 0.5 as half that time, or $5 \times 10^6$ years. While this last timescale is not particularly important for characteristic paths in field and group environments, it is crucial for the densest environments, clusters. This is because the characteristic FRII jet feedback in cluster environments is responsible for both a SFR enhancement as well as a modification of the accretion state, from thin disk to advection dominated flow (Garofalo, Evans & Sambruna 2010). This implies a drop in the accretion rate. The theoretical boundary between thin disk and advection dominated accretion is at 0.01 the Eddington limit. As a result of this transition in accretion state while the system is in the process of spinning the BH down, the timescale to reach zero BH spin is increased. Since the accretion rate does not simply instantaneously drop to 0.01 the Eddington rate, the timescale to reach zero spin ends up in the range $10^7 - 10^9$ years. From Figure 2, we then associate SFR values with these times for cluster environments. In short, we have the data to build characteristic values of SFR and jet power as a function of time. We therefore produce a jet power vs. star formation rate plot also showing the times and the direction of time evolution along the paths for characteristic environment-dependent curves in Figures 3 and 4. Blue, red, and green objects represent characteristic locations at specific times for radio AGN triggered by mergers in clusters, groups, and fields, respectively. The blue, red, and green paths of Figure 4 represent characteristic paths for jetted AGN in clusters, groups, and fields with the arrows showing the evolution in time from the initial state of star formation along the radio quiet AGN line. All paths initially evolve to the right and downward as the star formation rate is enhanced while the black hole spin decreases (which makes the jet power decrease as well). Following the transition through zero spin, the tilt in accretion disk, and the emergence of a tilted jet as the spin increases, we have increasing star formation suppression along with increase in jet power. Clearly, no overall direct correlation is generated between jet power and star formation rate. This is the first time we provide a theoretical description of the jet power as a function of star formation rate from this paradigm.

We can also make predictions. Given the time evolution, objects in denser environments evolve more leftward from their initial location compared to other environments. This means that the redshift distribution of the objects in Jin et al (2025) should be higher for the objects on the right and lower for those on the left. This is true in an average sense only, given that initial paths should exist that are more rightward which would produce less leftward late time locations on the jet power-SFR plane. In addition to this, we can predict the excitation level for the objects of Jin et al (2025). This is because mergers lead to high excitation around accreting black holes that tend to evolve into low excitation over time, more so in rich environments (Garofalo, Evans & Sambruna 2010; Garofalo, Christian et al 2023). As a result, we predict that on average the excitation level of the sources will be lower for objects on the left and higher for those on the right. Because of the low excitation of these sources at late times, their accretion rates are low, and this means that the black hole spin will not reach high values during corotation because the time to spin it up is insufficient. This is why the jet power in corotation is markedly lower than during the counter-rotating phase. In field environments, on the other hand, the excitation level drops less on average given the weaker feedback, and this means that the jet from the black hole becomes suppressed (i.e. at high black hole spin the jet is suppressed in radiatively efficient disks – Neilsen & Lee 2009; Garofalo, Evans & Sambruna 2010; Ponti et al 2012). This explains the shorter green path in Figure 4.

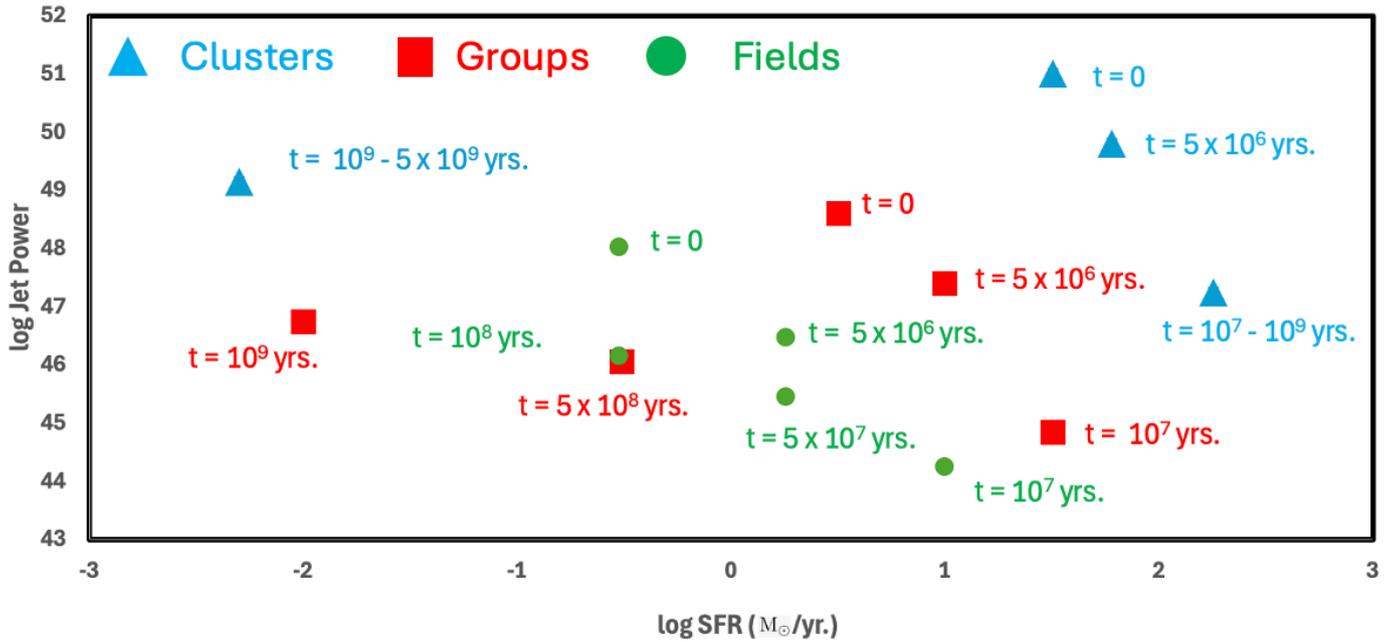

**Figure 3.** Characteristic location of radio AGN in the jet power versus SFR plane for cluster, groups, and field environments at characteristic times since their triggering. Jet power in units of erg s$^{-1}$ (B$_d$/10$^5$ Gauss)$^{-2}$.

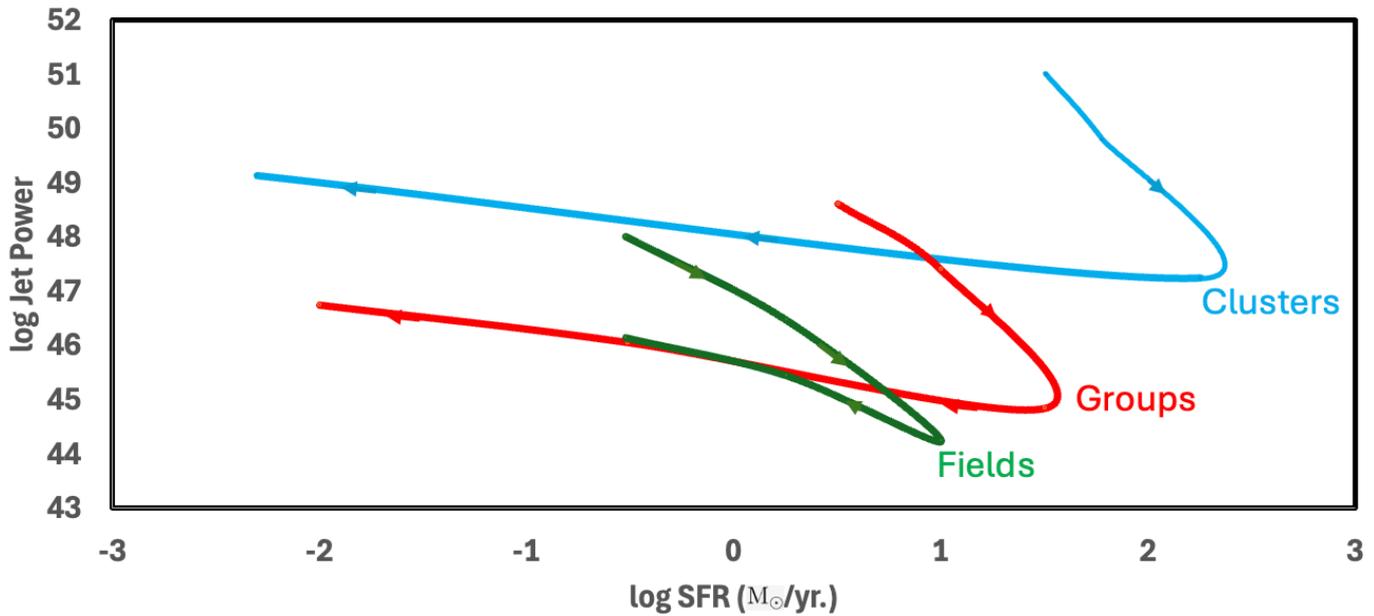

**Figure 4.** Jet power versus SFR with time direction for three characteristic curves for clusters, groups, and fields. Turning points for these curves represent transition from counter-rotation to corotation. Jet power in units of erg s$^{-1}$ (B$_d$/10$^5$ Gauss)$^{-2}$. This is our explanation for the lack of a correlation in the objects explored in Jin et al (2025).

Yamamoto et al 2024, have explored the spectral energy distributions (SEDs) of 7 high redshift radio galaxies by adopting a delayed star formation model as follows

$$dm_{stars}/dt = b(t/\tau) \exp(-t/\tau) \quad (3)$$

with t the time since star formation began, $\tau$ the peak time for star formation, and b an unspecified constant. From the SEDs, they have determined the best fit value of $\tau$ to be $10^8$ years. We modify the form of this expression to capture the putative initial conditions associated with when the active galaxy is triggered, which is that the star formation rate of the galaxy places it on the star formation main sequence and increases from that value until jet suppression occurs. In rich environments Singh et al (2021) prescribe peak star formation rates around 200 solar masses per year which fixes the form of the equation, and the constant as follows.

$dm_{stars}/dt = 500 (t/\tau) \exp(-t/\tau) + 35 \exp(-t/\tau)$.  (4)

We plot SFR versus time t in Figure 5 using the Yamamoto et al value of $\tau = 10^8$ years. In rich environments we know that the counter-rotation/prograde rotation boundary for black hole accretion occurs between $t = 10^7$ years and $t = 10^9$ years (Figure 3 blue data points). Because the jet power decreases as that boundary approaches, the ability of jet feedback to enhance the SFR must last less than $10^9$ years so the actual range for the SFR to increase must be between $10^7$ years and a few times $10^8$ years. The curve in Figure 5 is compatible with this as it peaks at $10^8$ years. The suppression of the SFR due to negative jet feedback in corotation must be compatible with the data of Comerford et al 2020 which shows that radio galaxies occupy a specific range in the SFR-stellar mass plane. In other words, we have a constraint on the number of stars formed and this needs to be captured by integrating equation (4). In short, the area under the curve of Figure 5 must satisfy $\log(M_{stars}/M_{solar}) > 10.6$. We estimate the area under the curve of Figure 5 to be about $8 \times 10^{10}$ solar masses which means $\log(M_{stars}/M_{solar}) = 10.9$ which is compatible with Figure 3 of Comerford et al 2020.

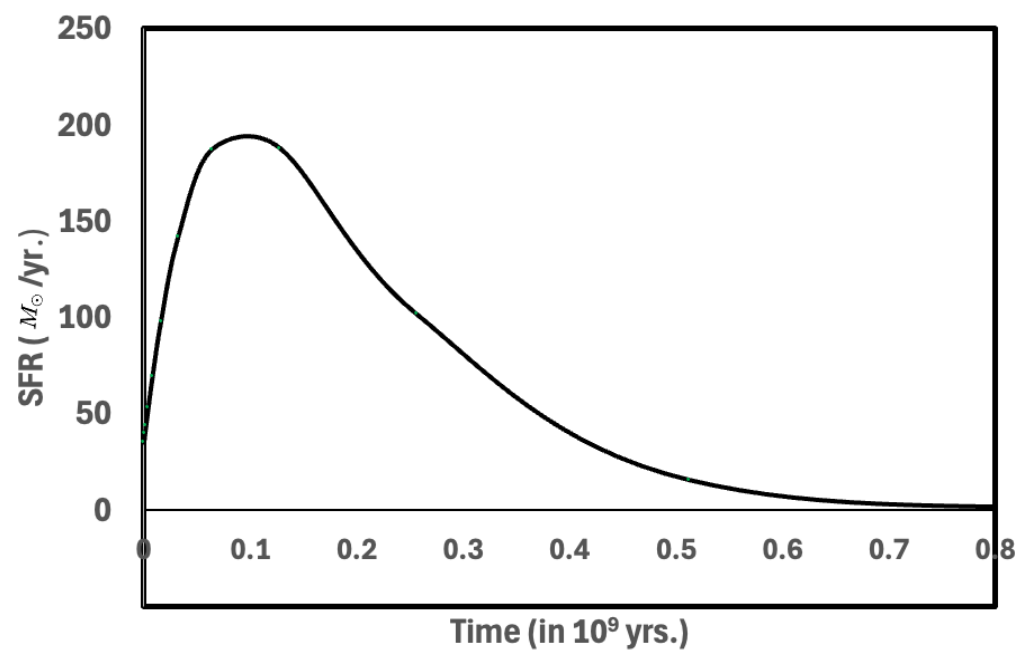

**Figure 5.** SFR versus time from equation 4.

## 3. Conclusions

Understanding whether black hole feedback is negative and/or positive in different circumstances and its connection to the formation and growth of galaxies is a key issue in high energy astrophysics. The evidence has been mixed across multiple parameters, but star formation rate is a crucial one. Over the past half decade, our understanding of the rate at which stars form in AGN has matured sufficiently to understand that many AGN are forming stars above the star formation main sequence, while others have rates that are well below that value. What kind of AGN feedback can enhance, and what kind can suppress, star formation? We have explored the data of Jin et al (2025) in the context of our model for black hole feedback and were able to show that radio AGN belong to a subclass of AGN that go through various phases that affect star formation but not in a unique and consistent manner. In their initial phase they enhance the star formation rate and later, they suppress it. While they enhance their star formation rates, their jet powers are decreasing and when they suppress their star formation rates, their jet power is increasing. We have shown how that feedback is different in a characteristic way depending on environment and we have made contact with the SEDs of high redshift radio galaxies. In closing we point out that the majority of mergers will trigger accreting black holes in corotation, which produces radio quiet AGN, and these objects are not part of our study. In fact, starburst galaxies likely belong to this subclass of newly triggered AGN. But their SFR would not benefit from an FRII jet enhancement.

# References

1. Bardeen J. M. and Petterson J. A. 1975 *ApJL* **195** L65
2. Cattaneo A., et al. 2009. Nature 460: 213–219
3. Comerford J. M., Negus J., Müller-Sánchez F. *et al.* 2020 *ApJ* **901** 159
4. Cresci, G. & Maiolino, R., 2018, Nature Astronomy, 2, 179
5. Croton, D. J., Springel, V., White, S. D. M., et al. 2006, MNRAS, 365, 11
6. Di Matteo, T., Springel, V., & Hernquist, L. 2005, Nature, 433, 604
7. Fabian, A.C. et al 2024, MNRAS, 535, 2173
8. Fabian, A. C. 2012, ARA&A, 50, 455
9. Ferrarese, L. & Merritt, D.A., 2000, ApJ, 539, L9
10. Garofalo, D., Evans, D.A., Sambruna, R.M., 2010, MNRAS, 406, 2
11. Garofalo, D., Joshi, R., Xiaolong, Y., Singh, C.B., North, M., Hopkins, M., 2021, ApJ, 889, 91
12. Garofalo, D., Christian, D.J., Hames, C., North, M., Thottam, K., Nazaroff, S., Eckelbarger, A., 2023, OJAp, 6, 27
13. Gultekin, K. et al. 2009, ApJ, 698, 198
14. Heckman, T. M., Kauffmann, G., Brinchmann, J., et al. 2004, ApJ, 613, 109
15. Jin, G., Kauffmann, G., Best, P. , Shenoy, S., Malek, K., 2025, A&A, 694, A309
16. Kalfountzou, E. et al 2012, MNRAS, 427, 2401
17. Kalfountzou E., Stevens J. A., Jarvis M. J. *et al.* 2014 MNRAS 442 1181
18. Kimmig, L.C. et al, 2025, ApJ, 979, 15
19. Kokorev, V. et al 2024, ApJ, 975, 78
20. Lowell, B., Jacquemin-Ide, J., Tchekhovskoy, A., Duncan, A. 2024, ApJ, 960, 82
21. Magorrian, J., Tremaine, S., Richstone, D., et al. 1998, AJ, 115, 2285
22. Neilsen, J. & Julia, L., 2009, Nature, 458, 481
23. Nesvadba, N.P.H. et al 2010, A&A, 521, A65
24. Nesvadba, N.P.H. et al 2006, ApJ, 650, 693
25. Ponti, G. et al, 2012, MNRAS, 422, L11
26. Joseph, P., Sreekumar, P., Stalin, C.S., Paul, K.T., Mondal, C., George, K.,, Matthew, B., 2022, MNRAS, 516, 2300
27. Reines, A. E. & Volonteri, M., 2015, ApJ, 813, 82
28. Sahu, N., Graham, A.W. & David, B.L., 2019, ApJ, 887, 10
29. Silk, J. & Rees, M.J., 1998, A&A, 331, L1
30. Silk, J., & Mamon, G.A., 2012, RAA, 12, 917
31. Silk, J., Begelman, M. C., Norman, C., Nusser, A., & Wyse, R. F. G. 2024, ApJ, 961, L39
32. Singh, C.B., Kulasiri, N., North, M., Garofalo, D., 2021, PASP, 133, 104101
33. Singh, C.B.; Williams, M.; Garofalo, D.; Rojas Castillo, L.; Taylor, L.; Harmon, E. *Universe* 2024, *10*, 319
34. Tchekhovskoy, A., McKinney, J., Narayan, R., 2012, Journal of Physics: Conf. Series 372, 012040
35. Yamamoto, Y. et al, 2024, arXiv:2411.19009v1
36. Zinn, P.-C. et al, 2013, ApJ, 774, 66